\documentclass[prl,twocolumn,superscriptaddress,floatfix]{revtex4}
\usepackage[dvips]{graphicx}
\usepackage{latexsym}
\usepackage{amsmath}
\usepackage{amsfonts}
\usepackage{amssymb}
\usepackage{bm}
\usepackage{txfonts}
\usepackage{xcolor}
\begin{document}

\newcommand{\fig}[2]{\includegraphics[width=#1]{#2}}
\newcommand{\kagome}{kagom\'e }
\newcommand{\bZ}{\mathbb{Z}}
\def\bk{{\mathbf{k}}}
\def\bK{{\mathbf{K}}}
\def\bM{{\mathbf{M}}}
\def\bkt{{\mathbf{\tilde{k}}}}
\def\bqt{{\mathbf{\tilde{q}}}}
\def\bR{{\mathbf{R}}}
\def\br{{\textbf{r}}}
\def\ba{{{\bm a}}}
\def\bX{{\mathbf{X}}}
\def\bq{{\mathbf{q}}}
\def\bQ{{\mathbf{Q}}}
\def\bg{{\mathbf{g}}}
\def\bG{{\mathbf{G}}}
\def\avs{{$A$V$_3$Sb$_5$}}
\def\kvs{{KV$_3$Sb$_5$}}
\def\cvs{{CsV$_3$Sb$_5$}}
\newcommand{\llangle}{{\langle\!\langle}}
\newcommand{\rrangle}{{\rangle\!\rangle}}

\title{Chern Fermi pocket, topological pair density wave, and charge-4e
 and charge-6e \\
superconductivity in \kagome superconductors}

\author{Sen Zhou}
\email{zhousen@itp.ac.cn}
\affiliation{CAS Key Laboratory of Theoretical Physics, Institute of Theoretical Physics, Chinese Academy of Sciences, Beijing 100190, China}
\affiliation{School of Physical Sciences \& CAS Center for Excellence in Topological Quantum Computation, University of Chinese Academy of Sciences, Beijing 100049, China}

\author{Ziqiang Wang}
\email{wangzi@bc.edu}
\affiliation{Department of Physics, Boston College, Chestnut Hill, MA 02467, USA}

\date{\today}

\begin{abstract}
The recent discovery of  novel charge density wave (CDW)  and pair density wave (PDW) in \kagome lattice superconductors $A$V$_3$Sb$_5$ ($A=$ K, Rb, Cs) hints at unexpected time-reversal symmetry breaking correlated and topological states whose physical origin and broader implications are not understood.
Here, we make conceptual advances toward a mechanism behind the striking observations and new predictions for novel macroscopic phase coherent quantum states.
We show that the metallic CDW state with circulating loop currents is a doped orbital Chern insulator near van Hove filling.
The emergent Chern Fermi pockets (CFPs) carry concentrated Berry curvature and orbital magnetic moment.
We find that the pairing of electrons on the CFPs leads to a superconducting state with an emergent vortex-antivortex lattice and the formation of a complex triple-$\bQ$ PDW.
A plethora of correlated and topological states emerge, including a never-before-encountered chiral topological PDW superconductor, a loop-current pseudogap phase, and vestigial charge-4$e$ and charge-6$e$ superconductivity in staged melting of the vortex-antivortex lattice and hexatic liquid crystal.
Our findings reveal previously unknown nature of the superconducting state of a current-carrying Chern metal, with broad implications for correlated and topological materials.
\end{abstract}
\maketitle

\noindent The field of transition-metal \kagome lattice materials has leapt forward with the discovery of superconductivity in a new family of vanadium-based \kagome metals {\avs} ($A=$ K, Rb, Cs) \cite{stephen-prm, stephen-prl}.
In contrast to the insulating \kagome compounds extensively studied for quantum spin liquids and doped Mott insulators \cite{yonglee, norman-rmp, palee}, {\avs} are nonmagnetic correlated metals with itinerant electrons traversing the geometrically frustrated \kagome lattice.
They are complementary to the (Fe, Co, Mn)-based itinerant \kagome magnets that exhibit rich correlated topological phenomena \cite{joe-nat, yin-nat, wsm-sci-morali, wsm-sci-liu, wsm-natphy-liu, yin-natphy}, but have remained nonsuperconducting at low temperatures.

All {\avs} undergo charge density wave (CDW) transitions below $T_{\rm cdw}\sim$ 78 -- 103 K and superconducting (SC) transitions below $T_c\sim$ 0.9 -- 2.5 K.
Both diagonal and off-diagonal long-range ordered states turn out to be highly intriguing and display a complex landscape of spontaneous symmetry breaking phases.
The CDW state has $2a_0\times2a_0$ charge order in the \kagome plane stacked along the $c$-axis \cite{cdw-natmat, xianhui-sc, cdw-nat, pdw-nat, binghai-prl, humiao-xray, stephen-xray, kvs-ilija}.
Despite the absence of magnetism, the CDW state shows evidence of spontaneous time-reversal symmetry (TRS)  breaking \cite{cdw-natmat, musr-1, musr-2, kerr} and exhibits giant anomalous Hall effect (AHE) \cite{g-ahe,xianhui-ahe}.
The CDW state also breaks rotation symmetry
\cite{cdw-natmat,cdw-nat,pdw-nat,kvs-ilija,haihuwen-twofold}.
Below $T_c$, a pair density wave (PDW) order was discovered \cite{pdw-nat}, where the finite-momentum pairing modulates the SC gap and coherence peaks with $\frac{4}{3}a_0 \times \frac{4}{3}a_0$ periodicity.
The primary PDW produces an unexpected pseudogap phase in striking comparison to the high-$T_c$ cuprates \cite{pdw-nat}.

The physical origin and implication of these remarkable observations are currently unknown. We study them here in order to understand quantum materials exhibiting both the intrinsic AHE and superconductivity.
We show that, close to the van Hove (vH) filling, the breathing \kagome lattice with circulating current gives rise to a loop-current Chern metal with Chern Fermi pockets (CFPs) carrying concentrated Berry curvature, which can account for the observed quantum oscillations \cite{stephen-xray,quantumosc-hechang,Shrestha-prb22} and the large intrinsic AHE \cite{g-ahe,xianhui-ahe}.
The SC state arising from the normal state Chern metal with loop currents turns out to be highly unconventional, since the supercurrents must circulate around spontaneously nucleated vortices.
We find that finite momentum pairing on the CFPs leads to novel roton PDW states with an emergent vortex-antivortex lattice.
Remarkably, due to the Fermi surface (FS) reconstruction in the $2a_0\times2a_0$ CDW state, the CFPs are connected by the wavevector ${\frac{3}{4}} {\bf Q}_{\rm Bragg}$, leading to the triple-$\bQ$ roton PDW with the observed ${\frac{4}{3}a_0 \times \frac{4}{3} a_0}$ periodicity \cite{pdw-nat}.

Superconductivity intertwined with broken spatial symmetries generates a landscape of extraordinary correlated and topological states that are relevant for current experiments.
We find a fully gapped chiral topological PDW superconductor protected by an induced uniform charge-$2e$ condensate.
The phases that straddle the SC ground state and the normal state Chern metal are intriguing.
Staged melting of the vortex-antivortex lattice and hexatic liquid crystal state by the proliferation of topological defects such as dislocations and disclinations produces spatially uniform charge-$4e$ superconductivity with orientational order and isotropic charge-$6e$ superconductivity.
Alternatively, destroying the SC phase coherence before restoring the crystalline symmetry leads to a loop-current pseudogap phase.



\section{Results}
\subsection{Breathing \kagome lattice}
\noindent In the {\avs} compounds, V atoms form an ideal \kagome lattice coordinated by Sb atoms and the alkali atoms ($A$) intercalate between the \kagome layers.
Very similar band structures are predicted across the series by the density functional theory (DFT) \cite{stephen-prm, binghai-prl}, in overall agreement with the measured band dispersions \cite{stephen-prl, yanzhang, sato, mingshi, comin, xinjiangzhou}.
There is an electron FS around the center of the hexagonal Brillouin zone (BZ) derived from the Sb $p_z$ orbital, while the actions of the low-energy V $d$-orbitals are located near the zone boundary, forming quasi-2D FS sheets close to the vH singularities.

To capture the most essential physics of a $d$ band crossing the Fermi level near the vH point, we consider the one-orbital model on the \kagome lattice as depicted in Fig.~1a with the lattice constant $a_0\equiv 1$.
The locations of the three sublattices in the unit cell at $\br$ are given by $\br_1=\br-{1\over2} \ba_3$, $\br_2=\br$, and $\br_3=\br +{1\over2} \ba_1$.
The tight-binding model for the corner-sharing up and down triangles of the \kagome lattice is written as,
\begin{equation}
H_\text{tb}=-\hspace{-0.15cm} \sum_{(\alpha\beta\gamma) \br} \left[t_{\alpha\beta}^u(\br) c_{\alpha \br}^\dagger c_{\beta \br} + t_{\alpha\beta}^d(\br) c_{\alpha \br}^\dagger c_{\beta \br- \ba_\gamma} + h.c. \right]-\mu \hat{n}_{\alpha \br},
\label{htb-realspace}
\end{equation}
where $c_{\alpha{\bf r}}^\dagger$ creates an electron on sublattice $\alpha$ in unit cell ${\bf r}$, $t_{\alpha\beta}^{u,d} ({\bf r})$ is the nearest-neighbor (nn) hopping in the up and down triangles between sublattices $\alpha$ and $\beta$, $\mu$ is the chemical potential, and $\hat{n}_{\alpha {\bf r}}=c_{\alpha{\bf r}}^\dagger c_{\alpha{\bf r}}$ is the density operator.
The sublattice indices run over $(\alpha,\beta,\gamma) =(1,2,3), (2,3,1), (3,1,2)$ and spin indices are left implicit.

\begin{figure}
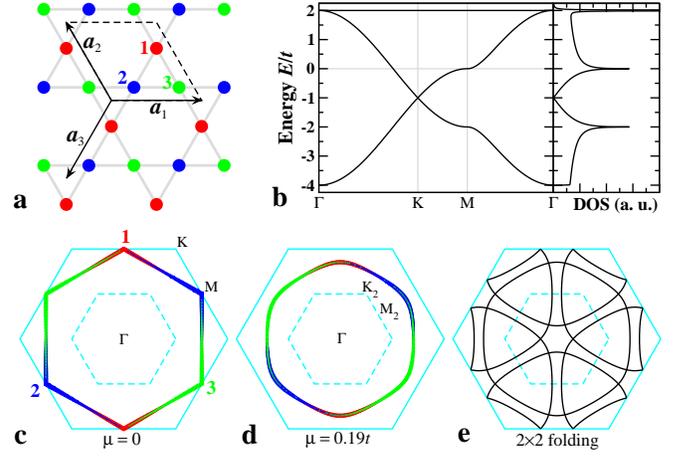

\begin{center}
\fig{3.4in}{fig1.eps}
\caption{Itinerant fermions on \kagome lattice.
\textbf{a} Lattice structure with {\em three} sublattices denoted by red ($1$), blue ($2$), and  green ($3$) circles. The two basis vectors defining the unit cell are $\ba_1=(1,0)$ and $\ba_2= (-{1\over2}, {\sqrt{3}\over2})$.
The reciprocal lattice vectors ${\bf b}_1={4\pi\over\sqrt{3}} ({\sqrt{3}\over2}, {1\over2})$ and ${\bf b}_2=
{4\pi\over\sqrt{3}} (0,-1)$.
The third direction follows $\ba_3=-\ba_1 -\ba_2$ and ${\bf b}_3=-{\bf b}_1-{\bf b}_2$.
\textbf{b} Band dispersion and DOS.
Fermi level at vH filling is marked by the grey line at $\mu=0$.
\textbf{c} FS at vH filling in the hexagonal BZ (solid cyan curves).
Thickness of colored lines and numbers display the sublattice contents.
The high symmetry points $\Gamma=(0,0)$, K$=(\pm{4\pi\over 3},0)$, $(\pm{2\pi\over 3},\pm{2\pi\over \sqrt{3}})$ and M$=(0, \pm{2\pi\over \sqrt{3}})$, $(\pm\pi,\pm{\pi\over \sqrt{3}})$.
\textbf{d} Sublattice resolved FS at $\mu=0.19t$ for electron doping above vH filling. Dashed cyan curves enclose the $2\times2$ reduced BZ with the resulting high symmetry points denoted by $\text{K}_2={1\over 2}\text{K}$ and $\text{M}_2={1\over 2}\text{M}$.
\textbf{e} FS in \textbf{d} folded by the $2\times2$ BZ boundary.
\label{fig:fig1}}
\end{center}
\vskip-0.5cm
\end{figure}

To reveal the unique geometry of the \kagome lattice, which plays a crucial role in the charge ordered state, we rewrite
\begin{equation}
H_\text{tb}=-\hspace{-0.15cm} \sum_{(\alpha\beta\gamma) \br} \left [t_{\alpha\beta}^+(\br) {\chi}_{\alpha\beta}^+ (\br)
+t_{\alpha\beta}^-(\br){\chi}_{\alpha\beta}^-(\br) +h.c. \right]-\mu \hat{n}_{\alpha \br},
\label{htb2}
\end{equation}
where $t_{\alpha\beta}^\pm={1\over2}(t_{\alpha\beta}^u\pm t_{\alpha\beta}^d)$ and the corresponding symmetric ($+$) and antisymmetric ($-$) nn bond operators,
\begin{equation}
\chi_{\alpha\beta}^\pm (\br)= c_{\alpha \br}^\dag c_{\beta \br} \pm c_{\alpha \br}^\dagger c_{\beta \br-\ba_\gamma}.
\label{bondops}
\end{equation}
For a \kagome lattice with uniform nn hopping, $t_{\alpha\beta}^+ (\br)=t$ and $t_{\alpha\beta}^- (\br)=0$.
The corresponding band dispersion and density of states (DOS) are shown in Fig. 1b.
A uniform $t_{\alpha\beta}^- (\br)= \delta\neq0$ produces an intra-cell breathing \kagome lattice \cite{nagaosa} without breaking lattice translation symmetry.
A spatially modulated hopping
$t_{\alpha\beta}^- (\br) =\delta \cos (\bQ\cdot \br)$ describes a breathing \kagome lattice with broken translation symmetry and a bond ordered CDW with wavevector $\bQ$.


\subsection{Complex CDW and Chern Fermi pockets}
\noindent At band filling $n_{\rm vH}=5/12$ or $\mu=0$, the sublattice-resolved FS is the hexagon connecting the vH singularity at the M points of the BZ in Fig.~1c.
Theoretical studies at this vH filling have demonstrated a rich set of instabilities toward correlated states due to the unique sublattice quantum interference effects \cite{thomale-prb, qhwang, thomale-prl}.
The extended Coulomb interactions $V$ can play a more important role than the local Hubbard $U$ and produce a $2a_0\times2a_0$ bond ordered CDW, which is indeed consistent with the triple-$\bQ$ CDW observed in the multi-orbital {\avs} where the Fermi level lies close to the $d$-bands vH singularities.
The DFT calculations \cite{binghai-prl} show that the $2a_0\times2a_0$ CDW is accompanied by the softening of the phonon breathing mode at the zone boundary M points of the \kagome lattice (Fig.~1c).
This corresponds to the triple-$\bQ$ breathing modulation of the antisymmetric bond in Eq.~(\ref{htb2}) and is described by the Hamiltonian
\begin{equation}
H_{\rm cdw}=\sum_{(\alpha\beta\gamma) \br} \rho_\gamma\cos(\bQ_{\rm c}^\gamma \cdot \br) \chi_{\alpha\beta}^-(\br) +h.c.,
\label{hcdw}
\end{equation}
where the CDW wavevector $\bQ_{\rm c}^\alpha ={1\over2} \bG^\alpha$, with the Bragg vector $\bG^{1,2,3}\equiv {\bf b}_{2,1,3}$.
The CDW maintains rotation and inversion symmetry, and exhibits the star-of-David (SD) and inverse-SD (tri-hexagonal)
bond configurations for  $\rho_\gamma >0$ and $\rho_\gamma <0$, respectively. Nevertheless, these bond ordered CDWs are real and do not break TRS.

Motivated by the conjecture of spontaneous TRS breaking orbital currents \cite{cdw-natmat}, a long-sought after quantum state also relevant for the pseudogap phase in the high-$T_c$ cuprates \cite{stagered-flux,ddw,varma} and the quantum anomalous Hall insulators \cite{haldane},
several theoretical studies have explored complex bond ordered triple-$\bQ$ CDWs supporting circulating current and plaquette flux \cite{chiralflux, thomaleneupert, balents, rhual}.
At the vH filling, the electron-electron interactions can drive a complex CDW state \cite{balents} and produce an orbital Chern insulator \cite{stagered-flux,rhual}.
It has also been mentioned that finite doping or imperfect nesting may lead to secondary orders from the residual Fermi surfaces, such as the unidirectional charge order and the triple-\textbf{Q} PDW \cite{rhual}.
Since the actual Fermi level is away from the vH points of the $d$-bands at M in all three {\avs} compounds \cite{binghai-prl,yanzhang, sato, mingshi, comin, xinjiangzhou},
it is important to study the complex bond ordered phase close to but not at vH filling.
As we show below, this results in a novel loop-current Chern metal with CFPs and an emergent primary PDW state detected in \cvs \cite{pdw-nat}.

We thus study the physics above vH filling ($n\gtrsim n_{\rm vH}$) by setting the symmetric hopping $t_{\alpha\beta}^+({\bf r})=t$ uniform as the energy unit and the chemical potential $\mu=0.19t$ in Eq.~(\ref{htb2}) with the sublattice resolved FS contour shown in Fig.~1d.
The triple-$\bQ$ complex CDW described by $H_{\rm cdw}$ in Eq.~(\ref{hcdw}) is generated by a complex $\rho_\gamma$, which amounts to a correlation-induced breathing modulation of the antisymmetric hopping $t_{\alpha\beta}^-(\br)= \rho_{\gamma}\cos(\bQ_{\rm c}^\gamma \cdot \br)$ in the single-orbital model in Eq.~(\ref{htb2}).
Besides being physically intuitive, we find that such a complex CDW state with equal amplitudes in the 3$\bQ$ directions $\rho_\gamma=\rho$ can be realized as the self-consistent mean field ground state driven by nn ($V_1$) and next-nn ($V_2$) Coulomb interactions (See Methods).
Diagonalizing $H=H_\text{tb} +H_{\rm cdw}$ with $\rho= (0.1+0.3i) (1,1,1)$,
we obtain the $2a_0\times2a_0$ complex CDW state shown in Fig.~2a characterized by three different current-carrying bonds.
Accumulating the link phases around the plaquettes produces four gauge-independent plaquette fluxes $\phi_{1,\dots,4}$ in the shaded $2\times2$ unit cell.
The flux is staggered since the total flux $\Phi= \phi_1 +3\phi_2 +2\phi_3 +6\phi_4$ is zero by symmetry. This orbital antiferromagnet breaks TRS but maintains inversion and C$_6$ rotation symmetry and gives rise to topological Chern bands on the \kagome lattice.

The band dispersion and the DOS are shown in Fig.~2b in the reduced BZ.
The Fermi level crosses the red Chern band with Chern number $C=3$ near the M$_2$ points, corresponding to electron doping of the Chern insulator at vH filling.
The CDW metal is therefore a doped orbital Chern insulator with emergent CFPs, as shown in Fig.~2c, residing along $\Gamma$-K and around M$_2$ along $\Gamma$-M direction due to the $2\times2$ band folding (Fig.~1e).
Each CFP has a volume $\sim0.56\%$ of the original BZ.
Remarkably, FS reconstruction and FS pockets of similar sizes have been detected by quantum oscillation experiments \cite{stephen-xray, quantumosc-hechang}.

\begin{figure}
\begin{center}
\fig{3.3in}{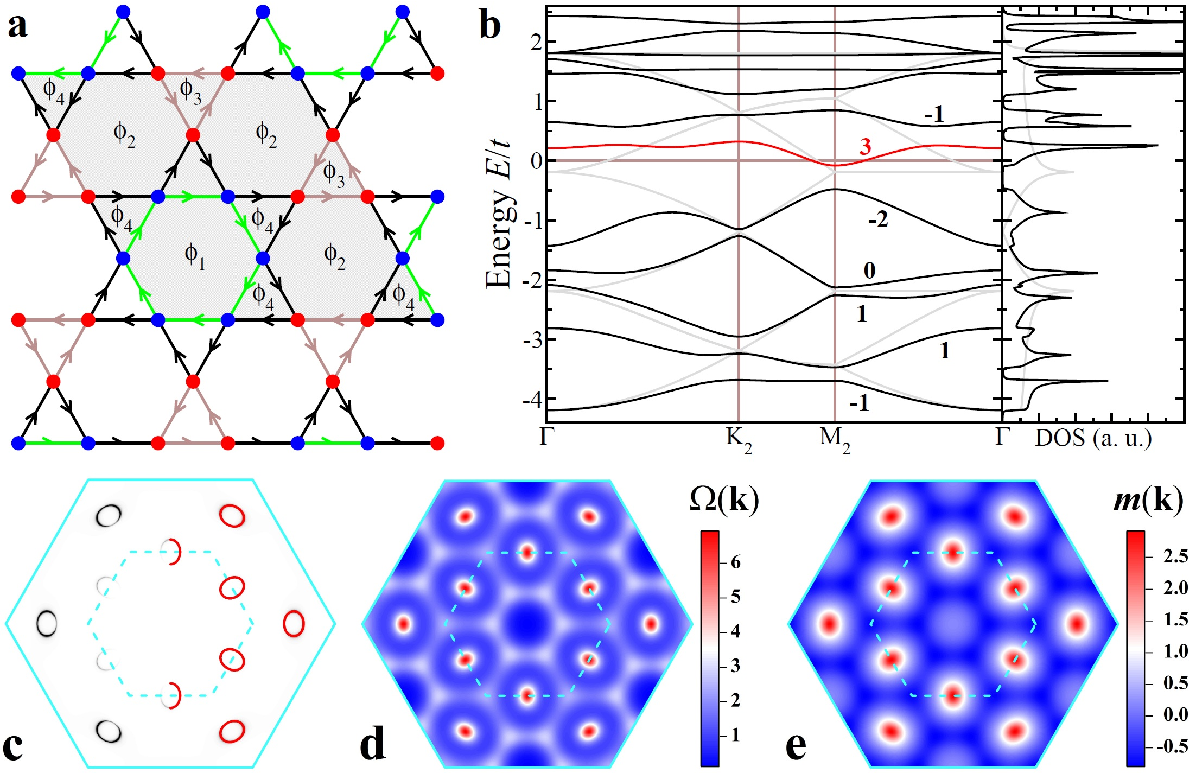}
\caption{Complex CDW and Chern Fermi pockets.
\textbf{a} Schematics of the $2a_0\times 2a_0$ CDW state. Electron density is 0.878 (0.810) on the red (blue) sites. Three complex hoppings on the black, green, and brown bonds are $0.499\pm 0.123i$, $0.345\pm 0.162i$, and $0.349\pm 0.110i$, respectively, giving rise to four independent plaquette fluxes $\phi_{1,\dots,4}=(0.840,-0.028,-0.291,0.014)\pi$.
The net flux through the unit cell (shaded grey) is zero.
\textbf{b} Band dispersion in the folded BZ and DOS.
Bands are marked by corresponding Chern numbers.
\textbf{c} Intensity plot of the DOS at the Fermi level, showing the spectral weight of the CFPs.
Locations of the pockets are superimposed as red solid ellipses on the right half of the BZ.
\textbf{d} Berry curvature distribution $\Omega(\bk)$ of the $C$=3 Chern band in \textbf{b}.
\textbf{e} Orbital magnetic moment distribution $m(\bk)$ of the $C$=3 Chern band in unit of $t a^2_0 e/ 2\hbar$.
\label{fig:fig2}}
\end{center}
\vskip-0.5cm
\end{figure}

To understand the intriguing properties of the CFPs, we calculate the Berry curvature and orbital magnetic moment (See Methods) of the hosting $C$=3 Chern band.
Fig.~2d shows that the Berry curvature in momentum space $\Omega(\bf k)$ concentrates heavily on the CFPs.
Remarkably, small orbits carrying large Berry curvature have indeed been detected in quantum oscillations \cite{stephen-xray, quantumosc-hechang}.
As a result, the CFPs contribute significantly to the AHE in addition to the fully occupied Chern bands in Fig.~2b: $\sigma_{xy}=2{e^2\over h}-2{e^2\over h}\int_{\circ}{d^2k\over 2\pi}\Omega({\bf k})\simeq 1.1{e^2\over h}$.
Using the $c$-axis lattice constant $c\simeq8.95$\AA\ \cite{stephen-prm}, the magnitude of the intrinsic anomalous Hall conductivity is estimated $\sigma_{xy}/c \simeq474 \Omega^{-1}{\rm cm}^{-1}$, which accounts for the observed large intrinsic AHE ($\sim500\Omega^{-1}{\rm cm}^{-1}$) \cite{g-ahe,xianhui-ahe}.
Moreover, the CFPs naturally lead to carrier density dependent intrinsic AHE observed by controlled gating in {\cvs} nanoflakes \cite{zhoujianhui}.

The calculated orbital magnetic moment $m(\bf k)$ is plotted in Fig. 2e for the $C=3$ Chern band hosting the CFPs in unit of $t a^2_0 e/ 2\hbar\simeq1.9\mu_B$ using the lattice constant $a_0\simeq5.4\AA$ and the hopping $t=0.5$eV, corresponding to a $3$eV bandwidth of the $d$-band.
The orbital moment, as large as $5\mu_B$, concentrates on the CFPs and couples to an applied magnetic field along the $c$-axis via the orbital Zeeman effect $-m(\bk) B_z$.
We thus expect strong and $\bk$-dependent field-induced responses in the electronic structure.
Moreover, the intrinsic thermodynamic orbital magnetization is calculated (Methods) to be $M \simeq -0.022\mu_B$ per V, which is large enough to be detectable by thermodynamic measurements.

\begin{figure*}
\begin{center}
\fig{7.0in}{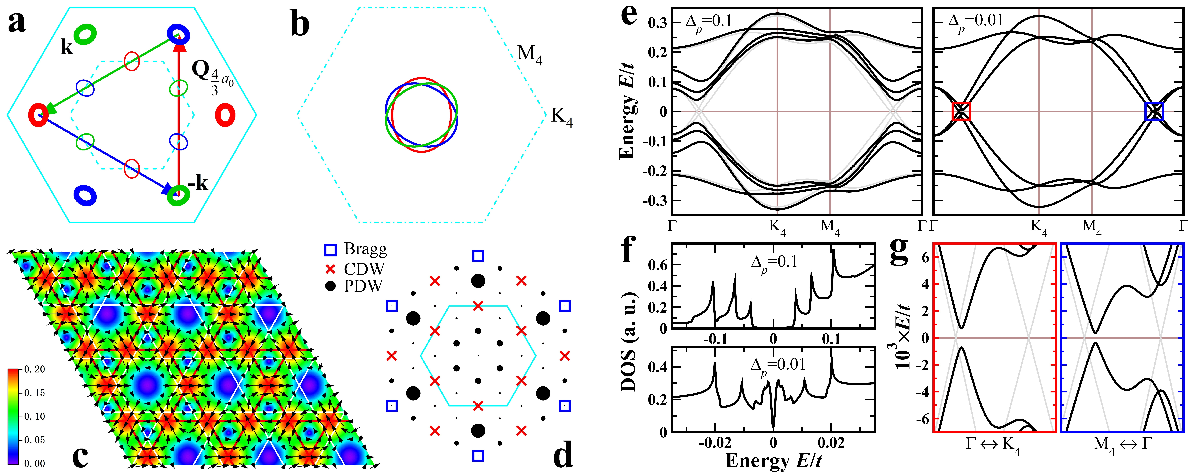}
\caption{3$\bQ$ PDW and gapped CFPs.
\textbf{a} Structure diagram of the CFPs. Pockets drawn in thick colored lines are connected by PDW wave vectors and carry most of the spectral weight (Fig.~\ref{fig:fig2}c).
\textbf{b} Folded elliptical pockets in the $4\times4$ reduced BZ.
\textbf{c} Spatial distribution of the ${4\over 3}a_0\times {4\over 3}a_0$ complex ${\rm PDW}_1$ order parameter for $\Delta_p=0.1t$, with amplitude represented by color intensity and phase denoted by arrows.
The emergent PDW \kagome lattice is shown by red lines, while the background \kagome lattice by white lines.
Vortex and antivortex at the center of the triangles and hexagons are visible by the phase winding indicated by the arrowed black lines.
\textbf{d} Fourier peaks under coexisting $2a_0\times2a_0$ CDW and ${4\over 3}a_0\times {4\over 3}a_0$ PDW$_1$.
\textbf{e} Low-energy BdG quasiparticle bands in PDW$_{1}$ for $\Delta_p=0.1t$ (left panel) and $0.01t$ (right panel) in the $4\times4$ folded BZ in \textbf{b}.
Grey lines correspond to $\Delta_p=0$.
\textbf{f} Low-energy DOS of PDW$_{1}$.
Top panel ($\Delta_p=0.1t$) shows three pairs of coherence peaks at the PDW gaps of the three low-energy bands.
Lower panel ($\Delta_p=0.01t$) shows a V-shaped SC gap.
A $k_BT=10^{-4}t$ thermal broadening is applied.
\textbf{g} Zoom-in of low-energy band dispersions inside red and blue boxes in the right panel of \textbf{e}, showing a full gap due to PDW induced uniform superconductivity.
\label{fig:fig3}}
\end{center}
\vskip-0.5cm
\end{figure*}


\subsection{Vortex-antivortex lattice and roton PDW}
\noindent The dynamically generated FS pockets are connected by well-defined momenta due to the band-folding.
Fig~3a displays the 12 color coded CFPs.
Pockets of the same color are related by the reciprocal lattice vectors of the $2a_0 \times 2a_0$ CDW $\bG_{2a_0}={1\over2} \bG$.
Those of different colors are, in contrast, connected by new wave vectors $\bQ_{4a_0}={1\over4} \bG$ and $\bQ_{{4\over 3} a_0}={3\over4} \bG$ in all 3\textbf{Q} directions, enabling new correlated quantum states that can coexist with the $2a_0\times2a_0$ CDW.

Note that neither $\bQ_{4a_0}$ nor $\bQ_{{4\over 3}a_0}$ leads to full nesting because of the ellipticity of the pockets (Fig. 3b).
Moreover, since the pockets are all electron-like, there is no instability in the particle-hole channel. Additional triple-$\bQ$ CDWs at these wave vectors cannot gap out the pockets (see Methods) and have not appeared experimentally, possibly because they are ineffective at lowering energy.
A striped $4a_0$ charge order was however observed by STM on the Sb surface in \cvs~\cite{cdw-nat,pdw-nat} which is allowed as a unidirectional single-$\bQ$ CDW in the Ginzburg-Landau (GL) analysis \cite{mcmillan}.

The path toward a stable correlated ground state can arise in the particle-particle channel through a PDW with finite momentum $\bQ_{\rm p}$ pairing of a spin-down electron at momentum $\bk$ and a spin-up electron at $-\bk+\bQ_{\rm p}$.
In contrast to a CDW, it does not require the connected CFPs to be electron and hole like to produce a large susceptibility.
The outer CFPs in Fig. 3a carry the majority of the spectral weight (Fig.~2c) and are more susceptible to a stabilizing PDW with ${4\over 3}a_0\times {4\over 3}a_0$ periodicity (Methods). The energetically favorable triple-$\bQ$ PDW therefore has the wave vector $\bQ_{\rm p}=\bQ_{{4\over 3}a_0}={3\over4} \bG$, which connects each of the six outer elliptical pockets in Fig.~3a with its two neighbors.
This corresponds precisely to the ${4\over 3}a_0\times {4\over 3} a_0$ periodicity of the observed PDW \cite{pdw-nat}.

Pairing of electrons on the CFPs with orbital currents has not been studied previously. The staggered normal state loop-current must be confined to loop-supercurrent modulating the phase and circulating the vortices in the SC state, which is thus inhomogeneous with an emergent vortex-antivortex lattice.
Since the microscopic pairing interaction is currently unknown, we consider the simplest onsite spin-singlet pairing,
\begin{equation}
H_{\rm pdw}=-\sum_{\alpha \br} 2\Delta_{\rm pdw}^\alpha (\br) c_{\alpha \downarrow}^\dagger(\br) c_{\alpha \uparrow}^\dagger(\br) + h.c.,
\label{hpdw}
\end{equation}
where $\Delta_{\rm pdw}^\alpha (\br)$ is the order parameter on the $\alpha$-sublattice in unit cell $\br$.
Incommensurate hexagonal PDWs have been studied
\cite{agterberg-hex}.
For commensurate PDWs in the \kagome superconductors, a symmetry analysis taking into account the \kagome lattice and the intrinsic $2a_0\times2a_0$ CDW is necessary.
The six basic PDW components carrying momenta $\pm \bQ_\text{p}^\eta$ and centered at $\br_0$ can be written as
$$
\Delta_{\pm \bQ^\eta_\text{p}} (\br)
=\Delta_p^{\pm\eta} e^{\pm i \bQ^\eta_\text{p} \cdot ( \br-\br_0)\pm i\phi_\eta},
$$
where $\eta=1,2,3$ denotes the three hexagonal directions and $2\phi_\eta$ is the relative phase between the $\pm \bQ_\text{p}^\eta$ modes.
Inversion and C$_6$ rotation symmetry requires that the amplitude and phase factor $\Delta_p^{\pm\eta}=\Delta_p e^{i\varphi_\eta}$, where $\varphi_\eta = \ell(\eta-1)2\pi/3$ with $\ell=0,\pm1$ stems from the eigenvectors of rotation.

In contrast to the incommensurate case, the center $\br_0$ of the commensurate PDW must be pinned to equivalent C$_6$ rotation centers of the loop-current CDW in Fig. 2a  (see Methods).
A coherent superposition gives the allowed PDW states
\begin{equation}
\Delta_{\rm pdw}^\alpha (\br)=e^{i\theta} \Delta_p \sum_{\eta=1,2,3} e^{i\varphi_\eta} \cos[\bQ_{\rm p}^\eta \cdot (\br_\alpha-\br_0)+\phi_\eta],
\label{pdworder}
\end{equation}
where $\theta$ is the global SC phase.
Since there are $2\times2$ such $\br_0$ locations within each PDW unit cell that are connected by a $\pi$-phase shift, the symmetry of the PDW states described by Eq.~(\ref{pdworder}) is $U(1)\times {\mathbb Z}_2\times {\mathbb Z}_2$ and the phase $\varphi_\eta$ can be defined as modulo $\pi$ (see Methods).
Among the possible PDWs, the complex PDWs with $\ell=\pm1$ are consistent with the broken TRS of the loop-current CDW, and the states with $\phi_\eta=(0,0,0)$ are found to have a lower mean field energy, which will be denoted as ${\rm PDW_{\pm1}}$ hereafter.
The spatial distribution of the PDW$_1$ order parameter forms an emergent ${4\over 3}a_0\times {4\over 3}a_0$ \kagome lattice as shown in Fig.~3c.
The zeros of the order parameter located in the middle of the triangles and hexagons are the centers of the single-vortex and double-antivortex, respectively.
Such a vortex-antivortex lattice, and thus the complex triple-$\bQ$ PDW$_{\pm1}$, describes the conjectured roton-PDW \cite{pdw-nat}, based on the notion that a roton corresponds to a tightly bound vortex-antivortex pair.

Including the $H_{\rm pdw}$, the total Bogoliubov-de Gennes (BdG) Hamiltonian can be diagonalized, giving rise to the Fourier peaks in the CDW and PDW field distributions in Fig.~3d.
The BdG quasiparticle spectrum is shown in Fig.~3e.
The PDW gap is expected to open first at the FS crossings \cite{palee-ampere} of the CFPs in Fig.~3b.
For large enough PDW amplitudes, the CFPs are fully gapped, as shown in the left panel of Fig.~3e at $\Delta_p=0.1t$.
Three gapped quasiparticle bands, stemming from the three-colored CFPs, emerge with the minimum gap loci offset from the Fermi momenta.
The corresponding low-energy DOS in Fig.~3f (top panel) reveals particle-hole symmetric gaps and three pairs of PDW peaks in remarkable agreement with the DOS spectrum observed in the SC state of \cvs \cite{pdw-nat}.

For weaker amplitude $\Delta_p$, the PDW is expected to only gap out the crossings of the CFPs in Fig.~3b along $\Gamma$-K$_4$ and  $\Gamma$-M$_4$ directions, leaving gapless excitations on the residual FS sections.
Fig.~3e (right panel) shows the quasiparticle dispersion of PDW$_1$ for an order of magnitude smaller $\Delta_p=0.01t$. Surprisingly, the gapless PDW does not appear, and upon zooming in, a full excitation gap is still visible in Fig.~3g.
Since the gap developed over the residual FS is small, the low energy DOS (Fig.~3f, bottom panel) shows a V-shaped SC gap inside the PDW gap at very low temperatures, with gap sizes $\sim0.7$meV and $10$meV for $t=0.5$eV, in agreement with STM observations \cite{pdw-nat}.
The fully gapped PDW state hints at an incipient uniform SC discussed below.


\subsection{Intertwined CDW with orbital current}
\noindent A primary PDW induces coexisting and intertwined electronic orders.
The secondary order can become the primary vestigial order after the melting of the PDW by quantum and/or thermal fluctuations  \cite{agterberg-natphys, berg, agterberg-hex, palee-ampere, palee-pdw, agterberg-review}.
We find that the hexagonal symmetry $\sum_{\eta}\bQ_{\rm p}^\eta=0$ and the intrinsic complex CDW lead to a plethora of novel intertwined and vestigial states.
The induced CDWs originate from the bilinear products
$$
{\rho}_{2{\bf Q}_{\rm p}^\alpha}^s({\bf r}) \propto \Delta_{{\bf Q}_{\rm p}^\alpha}(\br)\Delta_{-{\bf Q}_{\rm p}^\alpha}^*(\br), \hspace{0.3cm} {\rho}_{{\bf Q}_{\rm p}^\alpha\pm{\bf Q}_{\rm p}^\beta}^s({\bf r}) \propto \Delta_{{\bf Q}_{\rm p}^\alpha} (\br)\Delta_{\mp{\bf Q}_{\rm p}^\beta}^*(\br),
$$
and carry wavevectors that coincide with those of the intrinsic CDW and primary PDW since $2\bQ_{\rm p}^\alpha = \bQ_{\rm c}^\alpha+ \bG^\alpha$, $\bQ_{\rm p}^\alpha +\bQ_{\rm p}^\beta =-\bQ_{\rm p}^\gamma$, and $\bQ_{\rm p}^\alpha -\bQ_{\rm p}^\beta =\bQ_{\rm p}^\gamma +\bG_{2a_0}^\alpha +\bG^\alpha$,
where ${\bf G}$ and ${\bf G}_{2a_0}$ are the reciprocal vectors of the original and the $2\times2$ CDW lattice.
Importantly, the complex ${\rm PDW}_{\pm1}$  generates a complex triple-$\bQ$ CDW at $\bQ_{\rm p}$ with nonzero angular momentum and orbital currents commensurate to those already present in the intrinsic loop-current CDW metal.


\subsection{Fluctuating PDW and pseudogap phase}
\noindent The PDW order parameter transforms under the phase change
$\Delta_{\pm \bQ_{\rm p}^\eta} \to e^{i\theta+i\varphi_\eta \pm i\Phi_\eta} \Delta_{\pm \bQ_{\rm p}^\eta}$, where
$\Phi_\eta =\bQ_p^\eta \cdot {\bm u}$ are the phonon modes associated with the displacement field ${\bm u}$ of the PDW vortex-antivortex lattice.
For a commensurate density wave order, it is well known that coupling to electrons leads to collective phase modes with an energy gap $\propto \Delta_p (\Delta_p/W)^{(M-2)/2}$, where $W$ is the bandwidth and $M$ the commensurability ratio \cite{leericeanderson}.
Due to the ${\mathbb Z}_2\times{\mathbb Z}_2$ commensurability locking, we have $M=2$, and the energy gap is on the order of the PDW gap, in contrast to the soft mode in the incommensurate case \cite{agterberg-hex}.
This can be the reason for the robustness of
the PDW observed in \cvs, having an energy gap around $5$ mV \cite{pdw-nat}.
When the proliferation of SC vortices in $\theta(\br)$ due to quantum and/or thermal fluctuations drives
$\langle\Delta_{\pm{\bf Q}_{\rm p}^\alpha}\rangle=0$, the induced CDWs  $\rho_{{\bf Q}_{\rm c,p}}^s$, unaffected by the SC vortices,
become the vestigial ordered states detectable directly by STM at the intrinsic CDW and PDW wave vectors.
The observed pseudogap phases \cite{pdw-nat}, both above $T_c$ and  in a strong magnetic field above $H_{c2}$, likely originate from such fluctuating PDW states.


\subsection{Intertwined BCS instability}
\noindent Intriguingly, the hexagonal symmetry allows a secondary uniform SC order \cite{agterberg-hex},
\begin{equation}
\Delta_{2e}^\alpha\propto \Delta^{\bm *}_{-\bQ^\alpha_\text{p}} (\br) \Delta_{\bQ^\beta_\text{p}} (\br) \Delta_{\bQ^\gamma_\text{p}} (\br) =\Delta_p^3 e^{i(\theta-2\varphi_\alpha)},
\label{charge2e}
\end{equation}
where $\sum_\eta\varphi_\eta=2\pi$ was used.
Accordingly, despite having only onsite pairing, a unique sublattice $d_{x^2-y^2}\pm id_{xy}$ superconductor with $\Delta_{2e}^\pm\propto \Delta_p^3 (1,e^{\pm i2\pi/3},e^{\mp i2\pi/3})$
arises and coexists with the primary roton ${\rm PDW}_{\pm1}$.
The induced uniform charge-2$e$ condensate ensures an intertwined BCS instability that fully gaps out the CFPs for arbitrarily small $\Delta_p$ as observed above.


\subsection{Charge-$\bm 4e$ and $\bm 6e$ superconductivity}
\noindent The most striking intertwined and vestigial orders are the charge-$4e$ and charge-$6e$ superconductivity described by the
spatially uniform higher-charge SC order parameters
\begin{equation}
\Delta_{4e}^\alpha\propto \Delta_{\bQ_{\rm p}^\alpha} \Delta_{-\bQ_{\rm p}^\alpha} =\Delta_p^2 e^{i2(\theta +\varphi_\alpha)}, \
\Delta_{6e}\propto \Delta_{\bQ_{\rm p}^\alpha} \Delta_{\bQ_{\rm p}^\beta} \Delta_{\bQ_{\rm p}^\gamma} =\Delta_p^3 e^{i3\theta}.
\label{charge6e}
\end{equation}
Decoupled from the translational displacement field $\Phi_\eta$ of the PDW, they describe coexisting condensates of four-electron (two Cooper pairs) bound states in $d\pm id$ symmetry and six-electron (three Cooper pairs) bound states in isotropic $s$-wave symmetry, and raise the potential for realizing vestigial higher-charge superconductivity \cite{agterberg-natphys, berg, agterberg-hex, liangfu4e, hongyao4e}.
Deep in the roton PDW state, the tightly bound vortex  pairs form the hexagonal solid depicted in Fig.~3c. The vestigial order can thus be described by the celebrated Kosterlitz-Thouless-Halperin-Nelson-Young (KTHNY) theory for the staged melting of a vortex-antivortex lattice \cite{kt,nh,young,lubensky}.
In the KTHNY theory, the proliferation of topological defects, i.e., translational dislocations in $\Phi_\eta$ drives $\langle e^{i\Phi_\alpha}\rangle=0$ and destroys the PDW order $\langle\Delta_{\pm \bQ_{\rm p}^\alpha} \rangle =0$, as the vortex-antivortex lattice loses translation order and melts into a liquid crystal where the vestigial $\Delta_{2e}$,  $\Delta_{4e}$, and $\Delta_{6e}$ prevail as the primary orders.

Note that, analogous to a nematic, this liquid crystal possesses discrete orientational order and is termed a hexatic \cite{nh}.
This can be seen in the $e^{\pm i2\varphi_\alpha}$ phase factors of $\Delta_{2e}$ and $\Delta_{4e}$ in Eqs~(\ref{charge2e}-\ref{charge6e}).
Thus this intermediate phase is a hexatic SC and the $\varphi_\alpha$ are related to the ordering directors. The hexatic order parameter is defined by $\Psi_6(\br)=e^{i6\varphi(\br)}$, where $\varphi(\br) ={1\over2}(\partial_x u_y-\partial_yu_x)$ measures the orientation of a neighboring vortex bond relative to a fixed axis under the displacement field \cite{nh}.
The hexatic SC is described by an effective low-energy Hamiltonian in the long-wavelength limit ${\cal H}/k_BT ={\rho_s\over2} ({\bm \nabla}\theta)^2+{K_A\over2} ({\bm \nabla}\varphi)^2 -g_6\cos(6\varphi)$ where $\rho_s$ is the superfluid stiffness, $K_A$ the Frank constant \cite{nh} for the hexatic stiffness, and $g_6$ the coupling constant for the six-fold anisotropy.
The charge $2e$, $4e$ and $6e$ fields exhibit power-law correlations $G_{2,4,6}\sim r^{-(\eta_s+4\eta_A)}, r^{-4(\eta_s+\eta_A)}, r^{-9\eta_s}$ with $\eta_s ={T/2\pi\rho_s}$ and $\eta_A={T/2\pi K_A}$ in the hexatic vortex-antivortex phase.

The topological excitations in the hexatic SC can be revealed by the single-valuedness of the charge-$2e$ order parameter $\Delta_{2e}^\alpha$ in Eq.~(\ref{charge2e}), i.e., along a path encircling a point defect $\oint d\theta-2\oint d\varphi_\alpha =n_\alpha\times2\pi$.
The equations imply independent SC $2\pi$-vortices in $\theta$ and $2\pi$-disclinations in $\varphi$ as well as $\pi$-single disclinations.
Surprisingly, the solutions also support fractional ${1\over3}$-vortices \cite{agterberg-hex} with a phase winding of $2\pi/3$ bound to ${1\over3}$-disclinations.
The phase structure of such logarithmically interacting topological excitations is well-understood in the KTHNY theory by the vector Coulomb gases \cite{kt,nh,young,agterberg-hex}, which can be described by an effective sine-Gordon field theory ${\cal H}_{\rm dual}$ for the dual fields of $\theta$ and $\varphi$ \cite{berg,hongyao4e}.
In the region $K_A< \rho_s$, the hexatic undergoes a second stage melting with the proliferation of double disclinations into an isotropic liquid, where the correlation of $\Psi_6(\br)$ decays exponentially and vortices are free.
As a result, independent of the details of the transition, charge-$2e$ and $4e$ condensates are destroyed by the exponential decay correlation of the $e^{\pm i2\varphi_\alpha}$ phase factors
in the isotropic liquid, $\langle\Delta_{2e}\rangle =\langle\Delta_{4e}\rangle=0$.
Intriguingly, since the charge-$6e$ order $\Delta_{6e}$ in Eq.~(\ref{charge6e}) does not couple to $\varphi$ and is unaffected by disclinations, its correlation remains algebraic. This is a novel SC phase with a charge-$6e$ condensate and confined fractional ${1\over3}$-vortices.
The transition from the charge-$6e$ SC to the normal state proceeds by the unbinding and proliferation of fractional $1/3$-vortices and $1/3$-disclinations.
While the detailed phase diagram deserves future studies, our findings suggest that the \kagome superconductors offer a promising platform for realizing higher-charge superconductivity.

Remarkably, charge-4$e$ and charge-6$e$ Little-Parks magnetoresistance oscillations have been observed very recently in mesoscopic ring structures fabricated using {\cvs} flakes \cite{jianwang}.
It is conceivable that making the ring devices introduces significant lattice defects and strain fields that weakened and enabled the melting of the PDW before the SC phase coherence.
Candidate charge-$4e$ and charge-$6e$ SC states are evidenced by the sequential change in the flux quantization from $hc/2e$ to $hc/4e$ and to robust $hc/6e$ over a widened fluctuation regime with increasing temperatures \cite{jianwang}.
Further theoretical and experimental studies are necessary to fully understand the possible realizations of higher-charge superconductivity.

\begin{figure}
\begin{center}
\fig{3.4in}{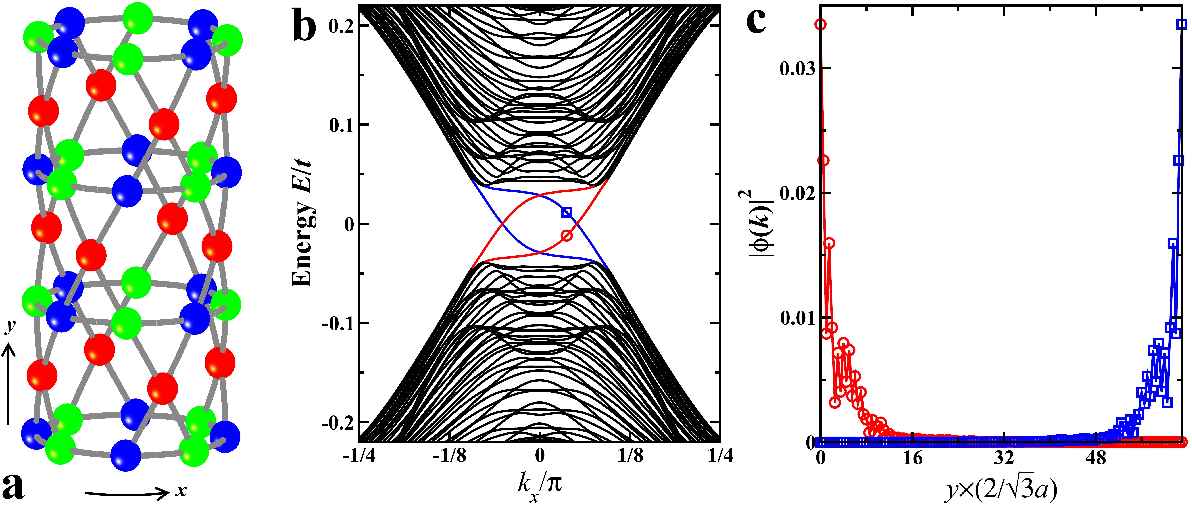}
\caption{Chiral topological PDW and chiral edge modes.
\textbf{a} Schematics of a cylindrical \kagome lattice with open boundaries in $y$-direction.
A single $4\times4$ supercell is shown.
Systems studied contain $L_x\times L_y$ supercells.
An infinite $L_x$ is used along the periodic direction in numerical calculations.
\textbf{b} Low-energy quasiparticle spectrum of the triple-$\bQ$ ${\rm PDW}_1$ at $\Delta_p=0.1t$, showing two in-gap CEMs (blue and red).
$L_y = 64$ with $4\times4$ supercell.
\textbf{c} Distribution along $y$-direction of the supercell averaged wavefunction amplitudes of the two in-gap CEMs indicated in \textbf{b} (red circle and blue square at $k_x=3\pi/80$), showing localization at the opposite edges illustrated in \textbf{\bf a}.
\label{fig:fig4}}
\end{center}
\vskip-0.5cm
\end{figure}


\subsection{\bf Chiral topological PDW}
\noindent The chiral PDW ground state with fully gapped CFPs carries an integer topological invariant ${\cal N}\in\bZ$ given by the total Chern number of the occupied BdG quasiparticle bands in Fig.~3e.
We obtain ${\cal N}=-2$ for ${\rm PDW}_{\pm1}$, indicating two charge neutral chiral edge modes (CEMs) at the sample boundary. Under inversion symmetric open boundaries in $y$-direction (Fig.~4a), the spectrum of the BdG Hamiltonian for PDW$_1$ is plotted in Fig.~4b, as a function of the momentum $k_x$ in the periodic $x$-direction.
Two CEMs of equal chirality inside the SC gap (red and blue in Fig.~4b) pairwise localize on the opposite edges (Fig.~4c).
We thus conclude that the triple-$\bQ$ complex PDWs are intrinsic chiral topological superconductors.


\section{Discussion}
\noindent In this work, we studied loop-current Chern metals with CFPs and found several thought-provoking correlated and topological states of pertinent interests in quantum condensed matter physics.
We proposed a novel mechanism for a class of PDW superconductors with loop-supercurrent circulating a vortex-antivortex lattice that emerge when the loop-current CDW metal enters the SC state.
It offers a plausible explanation for the ${4\over 3}a_0\times {4\over 3}a_0$ PDW observed in \cvs.
The \kagome superconductors also exhibit fascinating phenomena associated with rotational symmetry breaking.
The implications are discussed in Methods together with recent experimental findings.
The microscopic origin of the pairing interaction in the multiorbital {\avs} is beyond the scope of this work.
It can certainly be mediated by phonons and electron-phonon coupling \cite{binghai-prl}, given the weak local Coulomb repulsion and the absence of magnetism in \avs.
In view of the orbital current and CFPs in the normal state, the pairing interaction can also be mediated by electronic fluctuations such as loop-current fluctuations \cite{fdw, loopcurrent, palee-ampere} and Pomeranchuk fluctuations of the reconstructed FSs in the CDW state \cite{lin}.
While a primary uniform charge-$2e$ superconductor is unlikely to develop from the loop-current CDW, a complex roton PDW with a $2a_0\times2a_0$ vortex-antivortex lattice is not ruled out (Methods), which can be studied by microscopic model calculations in the future.
In addition to the \kagome superconductors \avs, our findings are relevant for quantum materials exhibiting both orbital-driven anomalous Hall effect and superconductivity, such as the superconducting moir\'e graphene.

\section{Acknowledgements}
We thank Binghai Yan, Kun Jiang, Jiangping Hu, Jiaxin Yin, Zahid Hasan, He Zhao, Ilija Zeljkovic, Hu Miao, Stephen Wilson, and especially Hui Chen and Hongjun Gao for valuable discussions.
S. Z. is supported by the National Key R\&D Program of China (Grant No. 2022YFA1403800), the Strategic Priority Research Program of CAS (Grant No. XDB28000000), and the National Natural Science Foundation of China (Grants No. 11974362 and No. 12047503). Z. W. is supported by the U.S. Department of Energy, Basic Energy Sciences (Grant No. DE-FG02-99ER45747) and by the Cottrell SEED Award No. 27856 from Research Corporation for Science Advancement. Numerical calculations were performed on the HPC Cluster of ITP-CAS. Z. W. thanks Aspen Center for Physics for hospitality and acknowledges the support of NSF Grant No. PHY-1066293.

\medskip
\noindent{\large \bf Methods}

\noindent{\bf Realization of complex CDW in ${\bm t}$-${\bm V}_{\bm 1}$-${\bm V}_{\bm 2}$ model.} 
The $t$-$V_1$-$V_2$ model consists of nn hopping $t$, nn Coulomb repulsion $V_1$, and next-nn Coulomb repulsion $V_2$,
\begin{equation}
H = - t \sum_{\langle i, j \rangle \sigma} (c^\dagger_{i\sigma} c_{j \sigma} + h.c.) +V_1 \sum_{\langle i, j \rangle} \hat{n}_i \hat{n}_j +V_2 \sum_{\llangle i, j \rrangle} \hat{n}_i \hat{n}_j,
\end{equation}
where $c^\dagger_{i \sigma}$ creates a spin-$\sigma$ electron on site $i$, and the density operator $\hat{n}_i=\sum_\sigma c^\dagger_{i\sigma} c_{i\sigma}$.
The Coulomb repulsions are decoupled in the density and bond channel by introducing $\hat{\chi}_{ij} =c^\dagger_{i\uparrow} c_{j\uparrow} +c^\dagger_{i\downarrow} c_{j\downarrow}$.
The mean-field Hamiltonian is,
\begin{align}
H_\text{MF} = -& t \sum_{\langle i, j \rangle \sigma} (c^\dagger_{i\sigma} c_{j \sigma} + h.c.) \\
-&V_1\sum_{\langle i,j \rangle} \left( \chi^*_{ij} \hat{\chi}_{ij} +h.c. - |\chi_{ij}|^2 -2n_i {\hat n}_j+n_i n_j \right) \nonumber \\
-&V_2 \sum_{\llangle i,j \rrangle} \left( \chi^*_{ij} \hat{\chi}_{ij} +h.c. -|\chi_{ij}|^2 -2n_i {\hat n}_j +n_i n_j\right), \nonumber
\end{align}
where the density $n_i=\langle {\hat n}_i\rangle$ and the complex bond order parameters $\chi_{ij}=\langle \hat{\chi}_{ij} \rangle$ are determined fully self-consistently.
We consider states that preserve $C_6$ rotation symmetry. The loop-current, when emerges, satisfies the continuity equation at the vertices of the \kagome lattice.

\begin{figure}
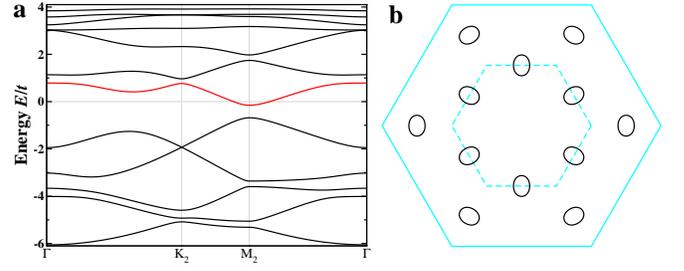

\begin{center}
\fig{3.4in}{figM1.eps}
\caption{Complex CDW in $t$-$V_1$-$V_2$ model.
Band dispersion (\textbf{a}) in the $2\times 2$ reduced BZ and the corresponding Chern FS pockets (\textbf{b}).
\label{figM1}}
\end{center}
\vskip-0.5cm
\end{figure}

At vH filling, we obtain an $2a_0\times2a_0$ triple-$\bQ$ complex CDW state, which is an orbital Chern insulator, driven by Coulomb repulsions $V_1=1.5t$ and $V_2=3t$.
Increasing the band filling to $n=5.064/12$, the self-consistently determined state is indeed a loop-current Chern metal with the band dispersion and the emergent CFPs shown in Fig. \ref{figM1}.
The pattern of the loop-current on the nn bonds is identical to the one shown in Fig. 2a in the main text, with the complex hoppings on black, green, and brown bonds given by $0.487\pm 0.071i$, $0.373\pm 0.074i$, and $0.367\pm 0.076i$, respectively.
A detailed study of the $t$-$V_1$-$V_2$ model is presented separately \cite{Dong22}.

\medskip
\noindent{\bf Berry curvature, anomalous Hall conductivity, orbital magnetic moment, and orbital magnetization.}
The Berry curvature of the $n$th quasiparticle band at momentum \textbf{k} is given by \cite{niu-rmp10}
\begin{align}
\Omega_n(\bk)&=i\langle \nabla_\bk u_{n\bk}| \times| \nabla_\bk u_{n\bk} \rangle \nonumber \\
&= i\left( \left\langle {\partial u_{n\bk} \over \partial k_x} \Big{|} {\partial u_{n\bk} \over \partial k_y} \right\rangle -\left\langle {\partial u_{n\bk} \over \partial k_y} \Big{|} {\partial u_{n\bk} \over \partial k_x} \right\rangle \right),
\end{align}
where $u_{n\bk}$ is the periodic part of the Bloch state wave function.
The Chern number of the corresponding band is obtained by integrating the Berry curvature over the BZ,
\begin{equation}
C_n = \int_{\rm BZ} {d\bk\over 2\pi} \Omega_n(\bk).
\end{equation}

The intrinsic contribution to the anomalous Hall conductivity in a 2D system is given by integrating the Berry curvature over all fully and partially occupied bands,
\begin{equation}
\sigma_{xy}=-
{e^2 \over h} \sum_n \int_{\rm BZ} {d\bk\over 2\pi} \Omega_n(\bk) f(\epsilon_{n\bk}),
\end{equation}
where $\epsilon_{n\bk}$ is the energy dispersion of the $n$th band relative to the chemical potential, and $f$ is the Fermi distribution function.
The calculated intrinsic anomalous Hall conductivity in the main text contains contributions from both the fully occupied Chern bands and the CFPs in the doped orbital Chern insulator.

\begin{figure}
\begin{center}
\fig{3.4in}{figM2.eps}
\caption{$4a_0\times 4a_0$ and ${4\over 3}a_0 \times {4\over 3} a_0$ CDW.
Low-energy bands (\textbf{a}) in the $4\times 4$ reduced BZ and corresponding FS pockets (\textbf{b}) in the presence of additional ${4a_0\times 4a_0}$ or ${{4\over 3}a_0 \times {4\over 3} a_0}$ CDW $(\rho=0.1)$. For comparison, band dispersion and FS pockets for the $2a_0\times 2a_0$ CDW without additional electronic order correspond to
the brown curves.
\label{figM2}}
\end{center}
\vskip-0.5cm
\end{figure}

\begin{figure*}[th]
\begin{center}
\fig{7.0in}{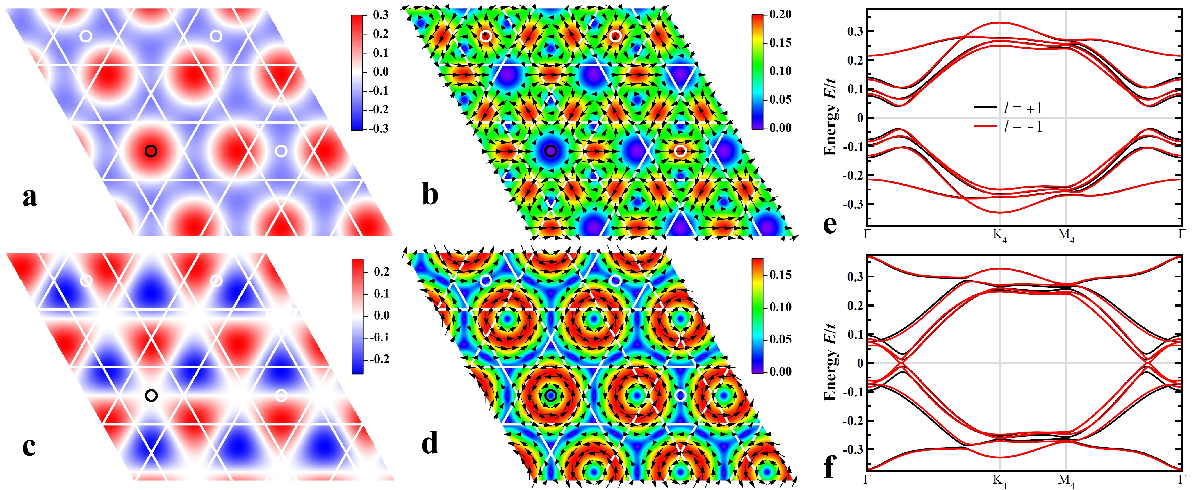}
\caption{Commensurate ${4\over 3}a_0 \times {4\over 3} a_0$ PDWs.
Spatial distribution of (\textbf{a}) PDW$_0$, (\textbf{b}) PDW$_1$, (\textbf{c}) PDW$'_0$, and (\textbf{d}) PDW$'_1$, evaluated with $\Delta_p = 0.1t$.
Small black and white circles mark the C$_6$ rotation and inversion centers of the loop-current CDW in Fig.~2a.
\textbf{e} Low-energy BdG quasiparticle bands for PDW$_{\pm 1}$.
\textbf{f} Low-energy BdG quasiparticle bands for PDW$'_{\pm 1}$.
In \textbf{a} and \textbf{c}, order parameters are real and represented by color intensity.
In \textbf{b} and \textbf{d}, order parameters are complex, with amplitude represented by color intensity and phase denoted by arrows.
\label{figM3}}
\end{center}
\vskip-0.8cm
\end{figure*}

The Berry curvature also induces an orbital magnetic moment of the $n$th band at momentum \textbf{k} \cite{niu-rmp10}
\begin{align}
{\bm m}_n(\bk)&=-i{e\over 2\hbar} \langle \nabla_\bk u_{n\bk}| \times (H_\bk - \epsilon_{n\bk}) |\nabla_\bk u_{n\bk} \rangle \nonumber \\
&= -i {e\over 2\hbar} \left( \left\langle {\partial u_{n\bk} \over \partial k_x} \Big{|} (H_\bk - \epsilon_{n\bk}) \Big{|} {\partial u_{n\bk} \over \partial k_y} \right\rangle -c.c \right) \hat{z}, \label{mk}
\end{align}
where $H_\bk$ is the Hamiltonian of the system.
Since the energy and wave vectors in this paper are measured in units of $t$ and $1/a_0$, respectively, the obtained values of  ${\bm m}(\bk)$ plotted in Fig.~2e is in units of $ta^2_0 e/2\hbar$.
The orbital magnetic moment ${\bm m}(\bk)$ couples to an out-of-plane magnetic field by the orbital Zeeman effect and offers an intriguing pathway for field-tuning of the band structure.
It is important to note that the orbital magnetic moment ${\bm m}(\bk)$ is different from the orbital magnetization ${\bm M}$ that enters the thermodynamic potential $G=E-\mu N-{\bm M}\cdot {\bm B}$ and is given by
\begin{equation}
{\bm M} =\sum_n \int_{\rm BZ} {d\bk\over (2\pi)^2} \left[ {\bm m}_n(\bk) -{e\over \hbar} \epsilon_{n\bk} \Omega_n(\bk) \right] f(\epsilon_{n\bk}).
\end{equation}
It contains contributions from both the fully occupied Chern bands and the CFPs in the doped orbital Chern insulator away from van Hove filling, and can be measured in thermodynamic experiments.

\medskip
\noindent{\bf $\bm{4a_0\times 4a_0}$ and $\bm{{4\over 3}a_0 \times {4\over 3} a_0}$ CDW.}
The $2a_0\times2a_0$ CDW order causes FS reconstruction and the formation of the CFPs connected by the dynamically generated wave vectors ${\bf Q}_{4a_0}$ or ${\bf Q}_{{4\over 3}a_0}$. Because the FS pockets are all electron-like, there is no instability in the particle-hole channel, making additional CDW formation at these wave vectors unfavorable.
To study their effect on the electronic structure, we consider additional ${4a_0\times 4a_0}$ or ${{4\over 3}a_0 \times {4\over 3} a_0}$ CDWs in the simplest form,
\begin{equation}
\widetilde{H}_{\rm cdw}=\sum_{\alpha \sigma \br} \widetilde{\rho}_\alpha (\br) c_{\alpha \sigma}^\dagger(\br)c_{\alpha\sigma}(\br),
\label{hcdw2}
\end{equation}
with an onsite CDW potential
\begin{equation}
\widetilde{\rho}_\alpha (\br) = \widetilde{\rho} \sum_\beta  \cos[\widetilde{\bQ}^\beta_\text{cdw} \cdot (\br_\alpha -\br_0)].
\end{equation}
Here, the ordering wave vectors $\widetilde{\bQ}_\text{cdw}$ is equal to ${\bf Q}_{4a_0}$ or ${\bf Q}_{{4\over 3}a_0}$, and $\br_0 = ({1\over 4}, {3\sqrt{3} \over 4})$ is located at the $C_6$/inversion center of the $2a_0 \times 2a_0$ CDW considered in the main text.

Diagonalizing $H_\text{tb} +H_\text{cdw} +\widetilde{H}_\text{cdw}$, the resulting low-energy bands and FS are shown in Fig. \ref{figM2} for a CDW potential  $\widetilde{\rho} =0.1$ in the $4\times4$ folded BZ.
The results in the absence of $\widetilde{H}_\text{cdw}$ are also plotted as brown curves for comparison.
It can be seen clearly that neither ${4a_0\times 4a_0}$ nor ${{4\over 3}a_0 \times {4\over 3} a_0}$ CDW can gap out the FS pockets.
Instead, they reshape the pockets into different sizes centered around $\Gamma$ in the $4\times 4$ reduced BZ.
As a result, the system cannot lower its energy significantly by forming such additional CDWs.

\medskip
\noindent{\bf Commensurate PDW order on kagom\'e lattice.}
In the main text, the general form of the triple-Q PDW order parameter respecting the C$_6$ rotation and inversion symmetries has been derived and given in Eq.~(\ref{pdworder}). We reproduced it here for the convince of discussion,
\begin{equation}
\Delta^\alpha_\text{pdw} (\br) = e^{i\theta} \sum_{\eta=1,2,3} e^{i\varphi_\eta}\Delta_p^\eta \cos \left[ \bQ_p^\eta \cdot (\br_\alpha -\br_0) +\phi_\eta \right], \label{Dpdw}
\end{equation}
where $\Delta_p^\eta=\Delta_p$, $\varphi_\eta = \ell (\eta-1)2\pi/3$ with  $\ell=0,\pm 1$, and $\br_0$ denotes the position of the commensurate PDW.
In the presence of the underlying \kagome lattice and parent $2a_0 \times 2a_0$ loop-current CDW, $\br_0$ must locate at the center of the C$_6$ rotation and inversion of the CDW shown in Fig.~2\textbf{a}, and equivalent locations under $2a_0 \times 2a_0$ translations.
Specifically,  $\br_0 =({1\over 4}, {3\sqrt{3} \over 4})+2n_1{\bf a}_1+2n_2{\bf a}_2$ with integers $n_1$ and $n_2$.
Additionally, the relative phase between $\pm \bQ_p^\eta$ modes, $\phi_\eta$, can only take the value of $(0, 0, 0)$ or $({\pi \over 2}, {\pi \over 2}, {\pi \over 2})$ that changes the three cosine functions to sine functions.

Hereinafter, we refer to the PDW with $\phi_\eta=(0, 0, 0)$ as PDW$_\ell$, and that with relative phase $({\pi \over 2}, {\pi \over 2}, {\pi \over 2})$ as PDW$'_\ell$.
The spatial distribution of the pairing order parameters of the $\frac{4}{3} a_0 \times \frac{4}{3} a_0$ triple-$\bQ$ PDWs are shown in Fig. \ref{figM3} for PDW$_0$ (\textbf{a}), PDW$_1$ (\textbf{b}), PDW$'_0$ (\textbf{c}), and PDW$^\prime_1$ (\textbf{d}).
PDW$_0$ and PDW$'_0$ are real and have circles of zeros and lines of zeros with maximum amplitudes forming triangular and honeycomb lattices, respectively.
The complex PDW$_1$ and PDW$^\prime_1$ have point zeros forming emergent vortex-antivortex lattices.

The symmetry of the PDW states described by Eq. (\ref{Dpdw}) is $U(1)\times \mathbb{Z}_2 \times \mathbb{Z}_2$, with $U(1)$ from the global SC phase.
The discrete symmetries come from the $2\times2$ equivalent locations of $\br_0$ in each $4\times4$ PDW unit cell, which are marked by small black and white circles in Fig. \ref{figM3}a-d.
Under a translation $\br_0\to \br_0+2\ba_{1,2}$, since $2\bQ_p^\eta \cdot \ba_{\eta'} =3\pi (1-\delta_{\eta \eta'}) $, the PDW components in Eq.~(\ref{Dpdw}) gain a $\pi$-phase shift in the cosine functions in two of the three hexagonal directions, while the remaining one is unaltered.
The $\mathbb{Z}_2 \times \mathbb{Z}_2$ symmetry arises because the transformed state can be brought back to the original one by a combination of $4a_0\times4a_0$ transition and inversion symmetry operations.
Equivalently, this symmetry can be understood as the BdG quasiparticle dispersions of the PDW being invariant under $\Delta_p^{\eta} \rightarrow -\Delta_p^{\eta}$ in Eq.~(\ref{Dpdw}).
As a result,
the relative phase $\varphi_\eta$ in Eq.~(\ref{Dpdw}) and Eq.~(\ref{pdworder}) in the main text can be defined as modulo $\pi$.

Because the parent loop-current Chern metal breaks TRS, only the complex triple-Q PDW states with $\ell= \pm 1$ need to be considered further.
Including the $H_\text{pdw}$ for the complex chiral PDW with $\ell=\pm 1$, the total BdG Hamiltonian can be diagonalized, giving rise to the quasiparticle spectrum shown in Fig.~\ref{figM3}e and \ref{figM3}f.
Clearly, compared to PDW$_{\pm 1}$, PDW$'_{\pm 1}$ of the same amplitude are much less effective at gapping out the quasiparticle bands and thus expected to be energetically unfavorable.
Calculating the mean-field state energies, defined as the expectation of the total BdG Hamiltonian $E_\text{PDW}=\langle H_\text{tb} +H_\text{cdw} +H_\text{pdw} \rangle$, we find that PDW$'_{\pm 1}$ have substantially higher energy than that of PDW$_{\pm 1}$, while the energy of PDW$_{1}$ is lower than that of PDW$_{-1}$ for the chosen normal state loop-current direction on the bonds.
Accordingly, we focused our study on PDW$_1$ in the main text.

Finally, we make connections to the incommensurate hexagonal PDW first studied in Ref.~\cite{agterberg-hex} in the effective continuum limit.
Without considering the underlying lattice and the parent commensurate CDW, the center of the PDW, $\br_0$, can slide freely along the three hexagonal directions.
This leads to two independent continuous and undetermined phases, $\phi'_1 = \bQ_p^1\cdot \br_0$ and $\phi'_2 = \bQ_p^2\cdot \br_0$, and a $\phi'_3= \bQ_p^3 \cdot \br_0 = -\phi'_1-\phi'_2$, corresponding to the two
Goldstone modes of the PDW states having $U(1)\times U(1) \times U(1)$ symmetry as discussed in Ref.~\cite{agterberg-hex}.
In this case, PDW$_0$, PDW$_{\pm 1}$, PDW$'_0$, and PDW$'_{\pm 1}$ recover $\Psi_\Delta$, $\Psi_\text{kag}$, $\Psi_\text{hc}$, and $\Psi_\text{hc,2}$ in Table I of Ref.~\cite{agterberg-hex}, correspondingly.

\begin{figure}
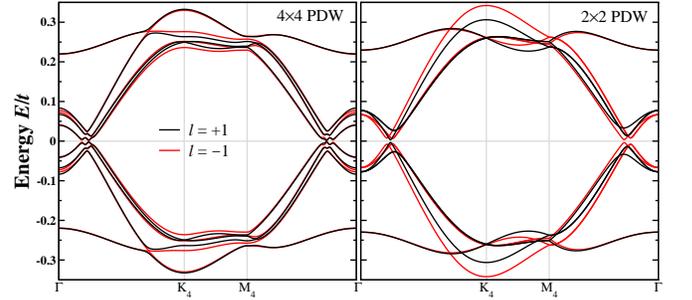

\begin{center}
\fig{3.4in}{figM4.eps}
\caption{$4a_0\times 4a_0$ and $2a_0\times 2a_0$ PDW.
Low-energy BdG quasiparticle bands for the 3\textbf{Q} PDW$_{\pm1}$ states with $\Delta_p = 0.1t$ along high-symmetry path in the $4\times 4$ reduced BZ.
The bands of the $2a_0\times 2a_0$ PDW are folded into the $4\times 4$ reduced BZ for the convenience of comparison.
\label{figM4}}
\end{center}
\vskip-0.6cm
\end{figure}

\medskip
\noindent{\bf $\bm{4a_0\times 4a_0}$ and $\bm{2a_0\times 2a_0}$ PDW.}
Fig. \ref{figM4} shows the low-energy BdG quasiparticle bands for the $4a_0\times 4a_0$ and $2a_0\times 2a_0$ PDW$_{\pm 1}$ states in the $4\times 4$ reduced BZ.
{Compared to those shown in Fig.~\ref{fig:fig3}\textbf{e} in the main text
for the ${4\over 3}a_0 \times {4\over 3} a_0$ PDW that pairs electrons on the outer FS pockets carrying the majority of the spectral weight (Fig. \ref{fig:fig3}\textbf{a}), the $4a_0 \times 4a_0$ and $2a_0 \times 2a_0$ PDW of the same amplitude are much less effective at gapping out the quasiparticle bands and thus not the energetically favorable PDW states.
This does not rule out the possible coexistence of these primary PDW states, which are all
capable of producing a smaller full gap due to the induced secondary uniform superconductivity.}

\medskip
\noindent{\bf Rotation and reflection symmetry breaking.}
In the original STM study of \kvs, the intensity of the six CDW peaks in the Fourier transform of the conductance map breaks {\em all} reflection symmetries and thus exhibits a handedness, which can be flipped by a magnetic field \cite{cdw-natmat}.
The existence of this chiral CDW is currently debated.
A later STM work \cite{kvs-ilija} finds that the handedness is absent, while a {\em single} reflection symmetry about one of the hexagonal directions, i.e. a reflection axis, remains and supports a rotational symmetry breaking CDW with $120^\circ$ oriented C$_2$ domains.
Moreover, the rotation symmetry is generally broken at high temperatures near the CDW transition \cite{hongli-recent}, independent of the $4a_0$ charge stripe order that appears around 50K in \cvs \cite{cdw-nat}, but not in \kvs \cite{cdw-natmat, kvs-ilija}.
However, the coherent electronic state with two-fold symmetry emerges only at low temperatures as indicated by the unidirectional quasiparticle interference patterns visible below about 35K \cite{hongli-recent}, which is consistent with the two-fold electronic state observed in NMR and elastoresistance measurements \cite{pottsnematic-cvs}.
These findings suggest that the rotation symmetry breaking down to  C$_2$ at $T_{\rm cdw}$ can be driven by the out of phase stacking of the in-plane $2a_0\times2a_0$ CDW with alternating SD and inverse-SD lattice distortions \cite{binghai-prl,balents,fernandes,miaohutheory}.

Most recently, all three $120^\circ$ oriented C$_2$ domains have been observed to onset at $T_{\rm cdw}$ by optical birefringence scanning probes in all three kagome metals {\avs}
\cite{liangwu}.
In each of the C$_2$ domains, TRS breaking is detected by optical Kerr rotation and circular dichroism \cite{liangwu}, which can be described by the complex CDW in Eq.~(\ref{hcdw}) with the real part of hopping in $\rho_\gamma$ along the reflection axis different from the other two directions.
The complex nonchiral phase factors \cite{rhual} in $\rho_\gamma$ produce a similar loop-current distribution.
Correspondingly, the triple-Q PDW can break the rotation symmetry with
the amplitude $\Delta_p^\eta$
in Eq.~(\ref{Dpdw}) taking a different value along the reflection axis.
Thus we expect the results discussed here to be applicable to {\avs} under weak rotation symmetry breaking.

\end{document}